\documentclass[12pt]{article}
\usepackage{epsfig}
\usepackage{graphics}
\usepackage{psfrag}
\usepackage{amssymb}

\newcommand{\beq}{\begin{equation}}
\newcommand{\eeq}{\end{equation}}
\newcommand{\bea}{\begin{eqnarray}}
\newcommand{\eea}{\end{eqnarray}}

\newcommand{\e}{\mbox{e}}
\renewcommand{\d}{\mbox{d}}
\newcommand{\g}{\gamma}

\newcommand{\lam}{\lambda}
\newcommand{\La}{\Lambda}
\renewcommand{\b}{\beta}
\renewcommand{\a}{\alpha}
\newcommand{\n}{\nu}

\newcommand{\Om}{\Omega}

\newcommand{\del}{\delta}

\newcommand{\kp}{\kappa}

\newcommand{\oh}{\frac{1}{2}}

\newcommand{\dg}{\dagger}

\newcommand{\tr}{\mathrm{tr}\,}
\newcommand{\Tr}{\mathrm{Tr}\,}
\newcommand{\ra}{\rangle}
\newcommand{\rra}{\right\ra}
\newcommand{\la}{\langle}
\newcommand{\lla}{\left\la}
\newcommand{\prt}{\partial}
\newcommand{\mi}{\!-\!}

\newcommand{\pl}{\!+\!}

\newcommand{\cD}{{\cal D}}

\newcommand{\cT}{{\cal T}}

\newcommand{\cO}{{\cal O}}

\newcommand{\tG}{{\tilde{G}}}

\newcommand{\tW}{{\tilde{W}}}
\newcommand{\tw}{{\tilde{w}}}

\newcommand{\tx}{{\tilde{x}}}

\newcommand{\tH}{{\tilde{H}}}

\newcommand{\hH}{{\hat{H}}}

\newcommand{\hG}{{\hat{G}}}

\newcommand{\hx}{{\hat{x}}}

\newcommand{\hp}{{\hat{p}}}

\newcommand{\bx}{{\bar{x}}}

\newcommand{\slt}{\sqrt{\lam} t}
\newcommand{\sla}{\sqrt{\lam}}
\newcommand{\sOm}{\sqrt{\Om}}

\newcommand{\vac}{|0\ra}
\newcommand{\cav}{\la 0 |}
\newcommand{\dll}{\frac{dl}{l}}

\newcommand{\rf}[1]{(\ref{#1})}

\begin{document}

\vspace{-36pt}

\begin{center}

{ \large \bf New aspects of two-dimensional quantum gravity}\footnote{Lectures presented at
the 49th Cracow School of Theoretical Physics, "Non-Perturbative Gravity and Quantum 
Chromodynamics",  Zakopane May 31-June 10, 2009. To appear in Acta Physica Polonica
B 40 (2009) 1001-1031. }

\vspace{12pt}

{\sl J.\ Ambj\o rn}$\,^{a,b}$, {\sl R.\ Loll}$\,^{b}$,
{\sl Y.\ Watabiki}$\,^{c}$, {\sl W.\ Westra}$\,^{d}$ and
{\sl S.\ Zohren}$\,^{e,f}$

\vspace{12pt}

{\footnotesize
$^a$~The Niels Bohr Institute, Copenhagen University\\
Blegdamsvej 17, DK-2100 Copenhagen \O , Denmark.\\
{ email: ambjorn@nbi.dk}\\

\vspace{6pt}

$^b$~Institute for Theoretical Physics, Utrecht University, \\
Leuvenlaan 4, NL-3584 CE Utrecht, The Netherlands.\\
{ email: loll@phys.uu.nl}\\

\vspace{6pt}

$^c$~Tokyo Institute of Technology,\\ 
Dept. of Physics, High Energy Theory Group,\\ 
2-12-1 Oh-okayama, Meguro-ku, Tokyo 152-8551, Japan\\
{email: watabiki@th.phys.titech.ac.jp}\\

\vspace{6pt}

$^d$~Department of Physics, University of Iceland,\\
Dunhaga 3, 107 Reykjavik, Iceland\\
{ email: wwestra@raunvis.hi.is}\\

\vspace{6pt}

$^e$~Mathematical Institute, Leiden University,\\
Niels Bohrweg 1, 2333 CA Leiden, The Netherlands\\
{email: zohren@math.leidenuniv.nl}\\

\vspace{6pt}

$^f$~Department of Statistics, Sao Paulo University,\\
Rua do Matao, 1010, 05508-090, Sao Paulo, Brazil

}

\end{center}

\vspace{6pt}

\begin{center}
{\bf Abstract}
\end{center}

Causal dynamical triangulations (CDT) can be used as a 
regularization of quantum gravity. In two dimensions
the theory can be solved anlytically, even before
the cut-off is removed and one can study in detail
how to take the continuum limit. 
We show how the CDT theory is related to Euclidean 2d
quantum gravity (Liouville quantum gravity), 
how it can be  generalized  and how 
this generalized CDT model has a string field theory representation
as well as a matrix model representation
of a new kind, and finally how it examplifies the 
possibility that time in quantum gravity might
be the stochastic time related to the branching 
of space into baby universes. 
%

\newpage

\section{introduction}\label{intro}

Presumably quantum gravity makes sense as an ordinary effective
quantum field theory at low energies. At high energies it is 
presently unclear how to view space-time. Is space-time an emergent 
low-energy structure as advocated in string theory, does 
it require a new concept of quantization as 
believed by people working on loop quantum gravity, or is 
quantum gravity ``just'' an almost standard quantum field 
theory with non-trivial (non-perturbative) ultraviolet behavior ?
As long as we do not know the answer we have an obligation
to pursue all avenues. 

The theory of quantum gravity which starts by providing an
utraviolet regularization in the form of a lattice theory, 
the lattice link length being the
(diffeomorphism invariant) UV cut-off, and
where in addition the lattice respects causality, is denoted CDT  
(causal dynamical triangulations). It is formulated in the spirit
of the last of the three approaches mentioned above: old ``boring''
quantum field theory with a non-trivial fixed point
\cite{weinberg}, \cite{reuteretc} . It allows
a rotation to Euclidean space-time, the action used being the 
Euclidean Regge action for the piecewise linear geometry represented
by the (now) Euclidean lattice (see \cite{ajl4d,blp} for details
of the Regge action in the CDT approach). 
The non-perturbative path integral is performed
by summing over lattices originating from Lorentzian lattices
with a causal stucture. Like in ordinary lattice field theories
we approach the contiuum by fine-tuning the bare coupling constants.
The rotation to Euclidean space-time allows the use of Monte Carlo 
simulations of the theory and in four-dimensional space-time, which 
for obvious reasons has our main interest, there exists a region
of coupling constant space  where
the infrared behavior of the universe seen by the computer is that 
of (Euclidean) de Sitter space-time \cite{agjl,emergence} (for  
a pedagodical review, see \cite{causality}). One might think this 
is a trivial result, but quite the contrary: in Euclidean space-time
the Einstein action is unbounded from below and the de Sitter 
solution is only a saddle point. Thus we are  clearly dealing
with an ``emergent'' property of the path integral, a genuine non-perturbative
effect arising from an interplay between the regularization and the 
path integral measure, and it can  only be valid in some 
limited region of the (bare) coupling constant space. This is 
precisely what is observed. One of the main questions to be answered
in such a lattice theory is what happens when the lattice 
spacing is taken to zero, i.e.\ when the cut-off is removed. 
Is it possible at all to remove the cut off, 
and if so what are the ultraviolet properties
of the theory ? Non-trivial UV properties have been observed 
\cite{semiclassical},
properties which have been reproduced by other ``field theoretical'' 
approaches to quantum gravity \cite{rg1,horava}. 
However, in the CDT-lattice approach it has so 
far been difficult to penetrate into the trans-Planckian region.
Active research is ongoing to achieve precisely this.

Numerical simulations are very useful for understanding whether a 
non-perturbatively defined quantum field theory has a chance
to make sense. Likewise it is useful for checking if 
certain conjectured non-perturbative features of the theory has a chance
of being true, and one can even discover new, unexpected phenomena.
In this way the numerical simulations work like experiments,
and this is the spirit in which the above mentioned simulations 
have been conducted. However, numerical simulations have their
limitations in the sense that they will never provide a proof
of the existence of a theory and it might be difficult in detail
to follow the way the continuum limit is approached since 
it requires larger and larger lattices. It is thus of interest
and importance to be able to study this in detail, even if
only in a toy model. Two-dimensional quantum gravity is such 
a toy model which has a surprising rich structure. Many of the 
intriguing questions in quantum gravity and in lattice quantum gravity
are still present in the two-dimensional theory.
What is nice about the two-dimensional theory 
is that it can be solved analytically even at the discretized level.

In the rest of this 
article we will discuss the two-dimensional CDT theory.
In Section \ref{cdt} we define the ``bare'' model. In Section \ref{cap1}
we generalize the model, allowing for local causality violations
and we discuss the generalized model's  relation to a specific matrix model
(Sec.\ \ref{matrix}). What is new is that the matrix 
model directly describes the continuum limit of the gravity theory.
In Sec.\ \ref{sft} we show how to formulate a complete
string field theory for the model and in Sec.\ 
\ref{matrix2} we show the equivalence to the matrix model 
defined in Sec. \ref{matrix}. The matrix model
allow us to define the theory
non-perturbatively and we show how to calculate
non-perturbatively (i.e.\ including the summation over all topologies)
certain observables. In Sec.\ \ref{other-m} we discuss relations
to other models.
In Section \ref{stochastic} we show that the string field theory
can be understood as a special kind of stochastic quantization 
of space, a phenomenon first noticed in the context of 2d Liouville 
quantum gravity in \cite{kawai}. Stochastic quantization defines  
a non-perturbative Hamiltonian. This is discussed in Sec.\ \ref{hamiltonian}.

\section{The CDT formalism}\label{cdt}

CDT stands in the tradition of 
\cite{teitelboim}, which advocated that in a gravitational path 
integral with the correct, Lorentzian signature of spacetime one should sum 
over causal geometries only. More specifically, we adopted this idea when it
became clear that attempts to formulate a {\it Euclidean} 
nonperturbative quantum gravity
theory run into trouble in space-time dimension $d$ larger than two.
Here we will discuss the implementation only when the space-time dimension
is two.

Thus we start from Lorentzian simplicial 
space-times with $d=2$ and insist
that only causally well-behaved geometries appear in the
(regularized) Lorentzian path integral. 
A crucial property of our explicit construction is that 
each of the configurations allows for a rotation to Euclidean signature
as mentioned above.
We rotate to a Euclidean regime in order to perform the sum over geometries
(and rotate back again afterward if needed).
We stress here that although the sum is performed over geometries 
with Euclidean signature, it is different from what one would 
obtain in a theory of quantum gravity based ab initio on Euclidean space-times.
The reason is that not all Euclidean geometries with a given topology are 
included in the ``causal'' sum since in general they have no correspondence 
to a causal Lorentzian geometry.
 
\begin{figure}[t]

\centerline{\scalebox{0.3}{\rotatebox{0}{\includegraphics{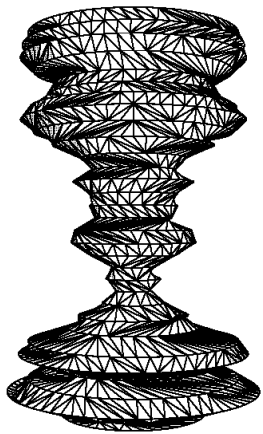}}}}
\caption{Piecewise linear space-time histories 1+1 dimensional quantum gravity}
\label{fig0}
\end{figure}

\begin{figure}[t]
\centerline{\scalebox{0.6}{\rotatebox{0}{\includegraphics{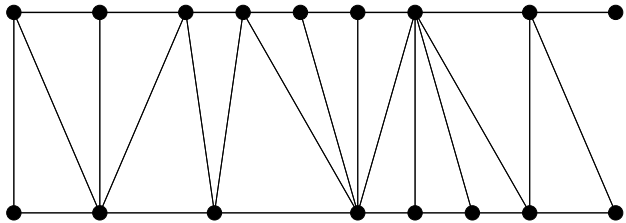}}}}
\caption{The propagation of a spatial slice from time $t$ to time $t+1$.
The ends of the strip should be joined to form a band with topology $S^1\times
[0,1]$.}
\label{2dminkowski}
\end{figure}

We refer to \cite{ajl4d} for a detailed description
of how to construct the class of piecewise linear geometries used 
in the Lorentzian path integral in higher dimensions. 
The most important assumption is the existence of 
a global proper-time foliation. This is illustrated
in Fig.\ \ref{fig0} in the case of two dimensions. We have a sum over 
two-geometries, ``stretching'' between two ``one-geometries'' separated 
a proper time $t$ and constructed from two-dimensional building blocks.
In Fig.\ \ref{2dminkowski} we have shown how to fill the two-dimensional
space-time between the space (with topology $S^1$) at time $t_n$ and 
time $t_{n+1}=t_n+a$ where $a$ denotes the lattice spacing. While we in the 
lattice model often use units where everything is measured 
in lattice length (i.e.\ the lattice links have length one),
we are of course interested in taking the limit $a \to 0$ to recover 
continuum physics.

In the path integral we will be summing over all possible ways 
to connect a given 1d ``triangulation'' at time $t_n$ and a given
1d triangulation at $t_{n+1}$ to a slab of 2d space-time as shown 
in Fig.\ \ref{2dminkowski},  
and in addition we will sum over all 1d ``triangulations''
of $S^1$ at times $t_n$. Thus we are demanding that the 
time-slicing is such that the topology of space does not change
when space ``evolves'' from time $t_n$ to time $t_{n+1}$. 

The Einstein-Hilbert action $S^{\rm EH}$ in two dimensions is 
almost trivial. According to the Gauss-Bonnett theorem the 
curvature term is a topological invariant and does not 
contribute to the equations of motion as long as the topology 
of space-time is unchanged. And even if the topology changes
the change in the curvature term is just a number, the change in
the Euler characteristic of the 2d surface. We will first ignore this
term, since we are first not allowing topology change.
Thus the (Euclidean) action simply consists of the cosmological term:  
\beq
S_E^{\rm EH}= \lam\int \d^2 x \;\sqrt{g}~~
\longrightarrow~~
 S_E^{\rm Regge}= 
\La N_2
\label{actshort}
\eeq 
where $N_2$ denotes the total number of triangles in
the two-dimensional triangulation. We denote the discretized
action {\it the Regge action} since it is a trivial example
of the natural action for piecewise linear geometries introduced 
by Regge \cite{regge}. The dimensionless lattice cosmological coupling
constant $\La$ will be related to the continuum cosmological 
coupling constant $\lam$ by an additive renormalization:
\beq\label{renor}
\La = \La_0 + \oh \lam\, a^2,
\eeq
the factor 1/2 being conventional.
The path integral or partition function for the CDT version of 
quantum gravity is now
\bea\label{2.1}
G^{(0)}_\lam(l_1,l_2;t) &=& \int \cD [g] \; 
\e^{-S_E^{\rm EH}[g]} ~~~\to \nonumber\\ 
G^{(0)}_\La(L_1,L_2,T) &=& 
\sum_{\cT} \frac{1}{C_\cT} \; \e^{-S_E^{{\rm Regge}}(\cT)},
\eea
where the summation is over all causal triangulations $\cT$ of the kind 
described above with a total of $T$ time steps, an ``entrance loop'' 
of length $l_1 = L_1 a$ and an ``exit loop'' of length $l_2= L_2 a$. 
The factor $1/C_\cT$ is a symmetry factor, given by the order of 
the automorphism group of the triangulation $\cT$. 

Our next task is to {\it evaluate} the sum over triangulations in \rf{2.1}, 
if possible, analytically. Surprisingly, it can be done \cite{al}. One 
can simply count the number of triangulations in a slice like the one 
show in Fig.\ \ref{2dminkowski} and from this the total number of 
triangulations in $T$ slices. As usual, when it comes to 
counting, it is often convenient to introduce the generating
function for the numbers one wants to count and first find
this function. In our model the generating function has a 
direct physical interpretation. We define
\beq\label{genfun}
\tG^{(0)}_\La(X_1,X_2;t) = \sum_{L_1,L_2} 
\e^{-X_1L_1}\e^{-X_2L_2} G^{(0)}_\La(L_1,L_2;T).
\eeq
Thus $\tG^{(0)}_\La(X_1,X_2;T)$ is the generating function of the 
numbers $G_\La^{(0)}(L_1,L_2;T)$ if we write $Z_1 = e^{-X_1}$, $Z_2=e^{-X_2}$.
But we can also view $X$ as a (bare) dimensionless boundary cosmological
constant, such that a boundary cosmological term $X \cdot L$ has 
been added to the action. In this way $\tG^{(0)}_\La(X_1,X_2;T)$ represents
the sum over triangulations where the lengths of the boundaries are 
allowed to fluctuate, the fluctuations controlled by the value $X$ 
of the boundary cosmological constant.  In general we expect, just
based on standard dimensional analysis, the boundary cosmological 
constants $X_i$ to be subjected to an additive renormalization when 
the continuum limit is approached. Like \rf{renor} we expect
\beq\label{renor1}
X = X_c +x\, a,
\eeq
where $x$ then denotes the continuum boundary cosmological constant,
and one, after renormalization, has the continuum boundary 
cosmological action $x\cdot l$.

We refer to \cite{al} for the explicit combinatorial arguments which 
allow us to find $\tG^{(0)}_\La(X_1,X_2;T)$. 
Let us just state the following results:
one can derive an exact iterative equation (using notation $Z=e^{-X}$,
$W=e^{-Y}$, $Q=e^{-\La}$)
\beq\label{cdt1}
\tG^{(0)}_\La(Z,W;T) = 
\frac{Q Z}{1- Q Z}\; \tG^{(0)}_\La\Big(\frac{Q}{1-Q Z},W;T-1\Big)
\eeq
This equation can be iterated and the solution written as 
\bea\label{cdt3}
\lefteqn{\tG^{(0)}_\La(Z,W;T) =}\\  
&& \frac{F^{2t}(1-F^2)^2\;ZW}{\Big[(1\!\!-\!ZF)\!-\!F^{2t+1}(F\!\!-\!Z)\Big]
\Big[(1\!\!-\!ZF)(1\!\!-\!WF)\!-\!F^{2T} (F\!\!-\!Z)(F\!\!-\!W)\Big]}~,
\nonumber
\eea
where $F$ is 
\beq\label{cdt4}
F=\frac{1-\sqrt{1-4Q^2}}{2Q}.
\eeq
These equations tell us that $Q_c = 1/2$ and that $Z_c = 1$ and
we can now take the continuum limit in \rf{cdt3} using $t=T \cdot a$ 
and find
\bea
\lefteqn{\tG^{(0)}_\lam(x,y;t) = 
\frac{4\lam\ \e^{-2\slt}}{(\sla+x)+\e^{-2\slt}(\sla-x)}}\hspace{1cm}\nonumber\\
&&\times \, \frac{1}{(\sla+x)(\sla+y)-\e^{-2\slt}(\sla-x)(\sla-y)}.
\label{cdt5}
\eea
Further, the continuum version of \rf{genfun}:
\beq\label{genfun1}
\tG^{(0)}_\lam(x_1,x_2;t) = \int_0^\infty \d l_1 \d l_2 \; \e^{-x_1l_1-x_2l_2} 
G^{(0)}_\lam(l_1,l_2;t).
\eeq
allows us to obtain:
\beq\label{cdt6}
G^{(0)}_\lam(l_1,l_2;t) = \frac{\e^{-[\coth \slt] \sla(l_1+l_2)}}{\sinh \slt}
\; \frac{\sqrt{\lam l_1 l_2}}{l_2}\; \; 
I_1\left(\frac{2\sqrt{\la l_1 l_2}}{\sinh \slt}\right), 
\eeq
where $I_1(x)$ is a modified Bessel function of the first kind.
Quite remarkable this expression was first obtained using 
entirely continuum reasoning by Nakayama \cite{nakayama}.

Finally, from \rf{renor} and \rf{renor1} 
one can now obtain the continuum limit of the iteration equation \rf{cdt1}:
\beq\label{cdt32} 
\frac{\prt}{\prt t} \tG^{(0)}_\lam(x,y;t) + \frac{\prt}{\prt x}
\Bigl[ (x^2-\lam) \tG^{(0)}_\lam(x,y;t) \Bigr]=0,
\eeq
This is a standard first order partial differential equation which 
should be solved with the boundary condition 
\beq\label{cdt78}
\tG^{(0)}_\lam(x,y;t=0)=\frac{1}{x+y}
\eeq
corresponding to
\beq\label{cdt79}
G^{(0)}_\lam(l_1,l_2;t=0)=\del(l_1-l_2).
\eeq
The solution is thus 
\beq\label{cdt33}
\tG^{(0)}_\lam(x,y;t) = \frac{\bar{x}^2(t;x)-\lam}{x^2-\lam}\; 
\frac{1}{\bar{x}(t;x)+y}, 
\eeq
where $\bar{x}(t;x)$ is the solution to the characteristic equation
\beq\label{cdt34}
\frac{\d \bar{x}}{\d t} = -(\bar{x}^2-\lam),~~~~\bar{x}(t=0)=x.
\eeq
It is readily seen that the solution is indeed given by \rf{cdt5}.
since we obtain
\beq\label{cdt35}
\bar{x}(t) = \sla \; 
\frac{(\sla+x)-\e^{-2\slt}(\sla-x)}{(\sla+x)+\e^{-2\slt}(\sla-x)}.
\eeq

If we interpret the propagator $G_\lam^{(0)}(l_1,l_2;t)$ as the matrix element
between two boundary states of a Hamiltonian evolution in 
``time'' $t$,
\beq\label{ham}
G^{(0)}_\lam(l_1,l_2;t)=<l_1|\e^{-{H}_0 t}|l_2>
\eeq 
we can, after an inverse Laplace transformation, read off the functional form
of the Hamiltonian operator $H_0$ from \rf{cdt32},
\beq\label{35b}
\tH_0(x) = \frac{\prt}{\prt x} \left(x^2-\lam) \right),~~~~~ {H}_0(l)=
 -l \frac{\partial^2}{\partial l^2}+\lam l .
\eeq

This end our short review of basic  2d-CDT.
We have here emphasized that all continuum results
can  be obtained by explicit solving the lattice model and 
taking the continuum limit simply by letting the lattice 
spacing $a \to 0$. The same will be true for the generalized
CDT model described below, but to make the presentation
more streamlined we will drop the explicit route via a lattice
and work directly in the continuum.

\section{Generalized CDT}\label{cap1}

It is natural the ask what happens if the strict requirement
of ``classical'' causality on each geometry appearing in
the path integral is relaxed. While causality is a resonable 
requirement as an outcome of a sensible physical theory, there
is no  compleling reason to impose it on each individual geometry
in the path integral, since these are not physical observables.
We used it, inspired by \cite{teitelboim}, as a guiding principle 
for obtaining a path integral which is different from the standard
Euclidean path integral, which was seemingly a necessity in 
higher than two space-time dimensions.

\begin{figure}[t]
\centerline{\scalebox{0.4}{\rotatebox{0}{\includegraphics{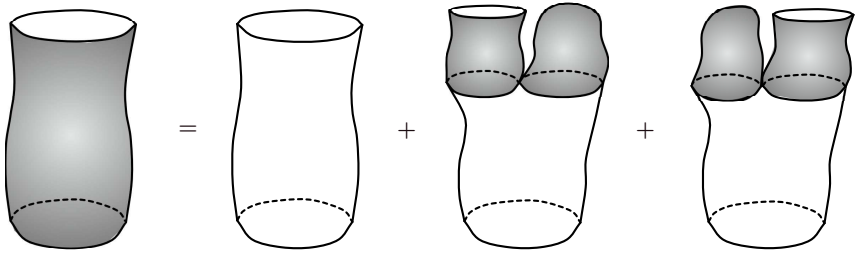}}}}
\caption{In all four graphs, 
the geodesic distance from the final to the initial 
loop is given by $t$. Differentiating
with respect to $t$ leads to eq.\ \rf{2.55}. Shaded parts of graphs represent
the full, $g_s$-dependent propagator and disc amplitude, and non-shaded 
parts the CDT propagator.}
\label{cap}
\end{figure}

In Fig.\ \ref{cap} we show what happens if we allow causality 
to be violated locally by allowing space to split in two at 
a certain time $t$, but we never allow the ``baby'' universe
which splits off to come back to the ``parent'' universe. 
The baby universe thus continues its life and is assumed
to vanish, shrink to nothing, at some later time. We now 
integrate over all such configurations in the path integral.
From the point of view of Euclidean space-time we are simply 
integrating over all space-times with the topology of a cylinder.
However, returning to the original Minkowskian picture it is clear
that at the point where space splits in two the light-cone is 
degenerate and one is violating causality in the strict local sense
that each space-time point should have a future and a past light-cone.
Similarly, when the baby universe ``ends'' its time evolution 
the light-cone structure is degenerate. These points thus have
a diffeomorphism invariant meaning in space-times with Lorentzian 
structure, and it makes sense to associated a coupling constant
$g_s$ with the process of space branching in two disconnected pieces.

The equation corresponding to Fig.\ \ref{cap} is \cite{alwz} 
\beq\label{2.55}
 \frac{\prt}{\prt t} \tG_{\lam,g_s}(x,y;t) = 
- \frac{\prt}{\prt x} \Big[\Big((x^2-\lam)+2 g_s\;W_{\lam,g_s}(x)\Big) 
\tG_{\lam,g_s}(x,y;t)\Big].
\eeq
$W_{\lam,g_s}(x)$ is denoted the disk amplitude with a fixed cosmological
constant $x$. It is related to the disk amplitude with a fixed boundary length by
\beq\label{l-disk}
\tW_{\lam,g_s}(x) = \int_0^\infty \d l \; \e^{-x l} W_{\lam,g_s}(l).
\eeq
It describes the  ``propagation'' of the a spatial universe until it vanishes in the 
vacuum. If we did not allow any spatial branching we would 
simply have
\beq\label{disk}
\tW^{(0)}_\lam (x) = \int_0^\infty \d t \; G^{(0)}_\lam(x,l=0;t) = 
\frac{1}{x+\sqrt{\lam}},
\eeq
where $G^{(0)}_\lam(x,l; t)$ denotes the Laplace transform 
of $G^{(0)}_\lam(l',l; t)$ with respect to $l'$.
From the composition rules for $G_{\lam,g_s}(l_1,l_2;t)$ it 
follows that it has (mass) dimension 1. Thus $G_{\lam,g_s}(x,l_2;t)$ is 
dimensionless  and it follows that the (mass) dimension 
of the coupling constant $g_s$ must be 3. In a discretized theory
it will appear as the dimensionless  combination $g_s a^3$, $a$ being
the lattice spacing, and one can show that the creation 
of more than one baby universe at a given time $t$ is suppressed by 
powers of $a$ (see \cite{alwz} for details). 
Thus we only need to consider the process shown in Fig.\ \ref{cap}. 
For a fixed cosmological constant $\lam$ and boundary
cosmological constants $x,y$ expressions like  $\tG_{\lam,g_s}(x,y;t)$
and $\tW_{\lam,g_s}(x)$ will have a power series expansion in the dimensionless
variable
\beq\label{kappa}
\kp = \frac{g_s}{\lam^{3/2}}
\eeq
and the radius of convergence is of order one. Thus the coupling 
constant $g_s$ indeed acts to tame the creation of baby universes 
and if $g_s$ exceeds this critical value eq.\ \rf{2.55} breaks down and 
is replaced by another equation corresponding to Liouville quantum gravity
with central change $c=0$. This was already observed in \cite{al}, and 
elaborated upon in \cite{ackl} where it was shown that 
one could indeed obtain the CDT from Liouville gravity by integrating 
out baby universes. In fact, in \cite{ackl} an amazing non-analytic map
between the two theories was discovered:
\beq\label{amaz}
\frac{x}{\sla} = \sqrt{\frac{2}{3}} \sqrt{ \frac{\tx}{\sqrt{\tilde{\lam}}}+1}
\eeq
where $\tilde{\lam}$ and  $\tx$ denote the cosmological and boundary cosmological 
constants in the purely Euclidean theory (Liouville quantum gravity with $c=0$).
This relation was rediscovered in \cite{seiberg2} where the CDT parameter
$x/\sla$ was called the uniformization parameter of
the algebraic surface which in \cite{seiberg2} was associated
to the semiclassical behavior of Liouville quantum gravity. CDT is the quantum theory
associated with the modified algebraic surface defined by the 
non-analytic mapping \rf{amaz}. Once the coupling constant $g_s$ is introduced
the CDT theory is charged, but in an analytic way, until the radius of convergence
for $\kp$ is reached. The detailed discussion of what happens when $\kp$ approaches
the radius of convergence can be found in \cite{alwwz}.

\begin{figure}[t]
\centerline{\scalebox{0.5}{\rotatebox{0}{\includegraphics{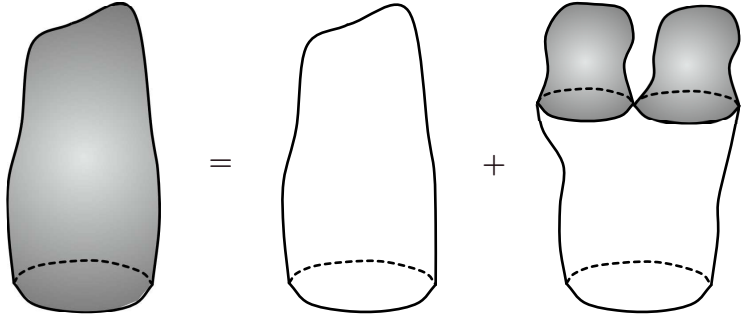}}}}
\caption[fig3]{{\small Graphical illustration of eq.\ \rf{3.2}. Shaded
parts represent the full disc amplitude, unshaded parts the CDT disc
amplitude and the CDT propagator. 
}}
\label{fig3}
\end{figure}

Differentiating the integral equation corresponding to Fig.\ \ref{cap}
fwith respect to the time $t$ one obtains \rf{2.55}.
The disc amplitude $W_{\lam,g_s}(x)$ is at this stage unknown. 
However, one has 
graphical representation for the disc amplitude 
shown in Fig.\ \ref{fig3}.
It translates into the equation \cite{alwz}
\bea\label{x3.2}
\lefteqn{\tW_{\lam,g_s} (x) = \tW_{\lam} ^{(0)}(x) +} \\
&&
g_s\int\limits_0^\infty \d t \int\limits_0^\infty \d l_1 \d l_2  \;
(l_1+l_2) G^{(0)}_{\lam}  (x,l_1+l_2;t) 
W_{\lam,g_s} (l_1)W_{\lam,g_s} (l_2) \nonumber
\eea
The superscript $(0)$ to indicate the CDT amplitudes without 
baby universe branching, calculated above. We assume
\beq
\tW_{\lam,g_s=0} (x)=\tW^{(0)}_\lam(x),\label{relabel}
\eeq 
and similarly for $G^{(0)}_{\lam,g_s}$. 
The integrations in \rf{x3.2} can be performed, yielding
\beq\label{x3.3}
\tW_{\lam,g_s}(x) = \frac{1}{x+\sla} +
\frac{g_s}{x^2-\lam}\Big( \tW_{\lam,g_s}^2(\sla)-\tW_{\lam,g_s}^2(x)\Big).
\eeq
Solving for $\tW_{\lam,g_s}(x)$ we find
\beq\label{x3.4}
\tW_{\lam,g_s}(x) = \frac{-(x^2-\lam) + \hat{W}_{\lam,g_s}(x)}{2g_s},
\eeq
where we have defined
\beq
\hat{W}_{\lam,g_s}(x) = \sqrt{(x^2-\lam)^2 +
4g_s\Big(g_s \tW^2_{\lam,g_s}(\sla) + x-\sla\Big)}.
\label{x3.4a}
\eeq
$\tW_{\lam,g_s}(x)$ is determined up to the value $\tW_{\lam,g_s}(\sla)$. 
We will now show that this value is fixed 
by consistency requirements of the quantum geometry. If we insert the 
solution \rf{x3.4} into eq.\ \rf{2.55} we obtain
\beq\label{x3.5}
 \frac{\prt}{\prt t} \tG_{\lam,g_s}(x,y;t) = 
- \frac{\prt}{\prt x} \Big[\hat{W}_{\lam,g_s}(x)\, \tG_{\lam,g_s}(x,y;t)\Big].
\eeq
In analogy with \rf{cdt32} and \rf{cdt33}, this is solved by
\beq\label{x3.6}
\tG_{\lam,g_s} (x,y;t) = 
\frac{\hat{W}_{\lam,g_s}(\bx(t,x))}{\hat{W}_{\lam,g_s}(x)} \; 
\frac{1}{\bx(t,x)+y},
\eeq
where $\bx(t,x)$ is the solution of the characteristic equation for \rf{x3.5},
the generalization of eq.\ \rf{cdt34}: 
\beq\label{x3.7}
\frac{\d \bx}{\d t} = -\hat{W}_{\lam,g_s}(\bx),~~~\bx(0,x)=x,
\eeq
such that
\beq\label{x3.8}
t = \int^x_{\bx(t)} \frac{\d y}{\hat{W}_{\lam,g_s}(y)}.
\eeq
Physically, we require that $t$ can take values from 0 to $\infty$, 
as opposed to just in a
finite interval. From expression \rf{x3.8} for $t$ this is 
only possible if the polynomial under the square root in the defining equation 
\rf{x3.4a} has a double zero, which fixes the function 
$\hat{W}_{\lam,g_s}(x)$ to
\beq\label{x3.9}
\hat{W}_{\lam,g_s}(x) = (x-\a)\sqrt{(x+\a)^2-2g_s/\a}=
\lam(\tx-u)\sqrt{(\tx+u)^2-2\kp},
\eeq
where 
\beq\label{3.9a}
x= \tx\sla,~~~\a = u\sla, ~~~u^3-u+\kp=0.
\eeq
In order to have a physically acceptable $\tW_{\lam,g_s}(x)$, 
one has to choose the solution to the third-order 
equation which is closest to 1 and the above statements about the 
expansion of $\tW_{\lam,g_s}(x)$ in a power series in $\kp$ follows
from \rf{x3.4}, \rf{x3.9} and \rf{3.9a}.

\section{The matrix model representation}\label{matrix}

The formulas \rf{x3.9} and \rf{x3.4} are standard
formulas for the resolvent of a Hermitean matrix model, calculated
to leading order in $N$, the size of the matrix. In fact  the following matrix model 
\beq\label{3.26}
Z(\lam,g_s) = \int d\phi \; 
\e^{-N\Tr V(\phi)},~~~~~
V(\phi) =  \frac{1}{g_s} \Big(\lam \phi -\frac{1}{3} \phi^3\Big)
\eeq
has a resolvent
\beq\label{resolv}
 \left\la \frac{1}{N}\;\Tr \left(\frac{1}{x-\phi}\right)\right\ra =
W_{\lam,g_s(x)}(x) + O(1/N^2),
\eeq
where $W_{\lam,g_s}(x)$ is given by \rf{x3.4}, and 
where the expectation value of a matrix expression $\cO(\phi)$ is 
defined as 
\beq\label{expect}
 \la \cO(\phi)\ra = 
\frac{1}{Z(\lam,g_s)}  \int d\phi \; 
\e^{-N\Tr V(\phi)} \; \cO(\phi).
\eeq

What is surpising here, compared to ``old'' matrix model approaches
to 2d Euclidean quantum gravity, is that the large $N$ limit reproduces
directly the continuum theory. No scaling limit has to be taken.
The situation is more like in the Kontsevich matrix model, which
directly describes continuum 2d gravity aspects. In fact the qubic 
potential is ``almost'' like the qubic potential in the Kontsevich 
matrix model, but the wold-sheet  interpretation is  different.

Can the above correspondance be made systematic in an 
large $N$ expansion and can the matrix model representation 
help us to a non-perturbative definition of generalized 2d
CDT gravity ? The answer is yes \cite{alwwz}.

First we have to formulate the CDT model {\it from first 
principles} such that 
we allow for baby universes to join the ``parent'' universe
again, i.e.\ we have to allow for 
topology changes of the 2d universe, and next we have to 
check if this generalization is correctly captured by the matrix 
model \rf{3.26} \cite{cdt-sft}

\section{CDT string field theory}\label{sft}

In quantum field theory particles can be created and annihilated
if the process does not violate any conservation law of the
theory. In string field theories one operates in the 
same way with operators which can create and annihilate strings.
From the 2d quantum gravity point of view we thus have a 
third-quantization of gravity: one-dimensional universes can 
be created and destroyed. In \cite{sft} such a formalism was
developed for non-critical strings (or 2d Euclidean quantum 
gravity). In \cite{cdt-sft} the formalism was applied to
2d cdt gravity leading to a string field theory or third quantization 
for CDT which allows us in principle to calculate any
amplitude involving creation and annihilation of universes.

Let us briefly review this formalism.
The starting point is the assumption of a vacuum from 
which universes can be created. We denote this state $\vac$ and
define creation and annihilation operators:
\beq\label{s1} 
[\Psi(l),\Psi^\dg(l')]=l\del(l-l'),~~~\Psi(l)\vac = \cav \Psi^\dg(l) =0. 
\eeq
The factor $l$ multiplying the delta-function is 
introduced for convenience, see \cite{cdt-sft} for a discussion.

Associated with the spatial universe we have a Hilbert space on the
positive half-line, and a corresponding scalar product (making $H_0(l)$
defined in eq.\ \rf{35b} hermitian) :
\beq\label{s4}
\la \psi_1 |\psi_2\ra = \int \frac{dl}{l} \; \psi_1^* (l) \psi_2(l).
\eeq
 
The introduction of the operators $\Psi(l)$ and $\Psi^\dg(l)$ in \rf{s1}
can be thought of as analogous to the standard second quantization 
in many-body theory. The single particle Hamiltonian $H_0$ defined by
\rf{35b} becomes in our case the ``single universe'' Hamiltonian. 
It has eigenfunctions
$\psi_n(l)$ with corresponding eigenvalues $e_n= 2n\sla$, $n=1,2,\ldots$:
\beq\label{s4a}
\psi_n(l) = l\, e^{-\sla l} p_{n-1}(l),~~~~~
H_0(l)\psi_n(l)= e_n \psi_n(l),
\eeq
where $p_{n-1}(l)$ is a polynomial of order $n\mi 1$.
Note that the disk amplitude $W^{(0)}_\lam(l)$, which is obtained from \rf{disk},
formally corresponds to $n=0$ in \rf{s4a}:
\beq\label{s4b}
W_\lam^{(0)}(l)=\e^{-\sla l},~~~~H_0(l) W_\lam^{(0)}(l)=0.
\eeq
This last equation can be viewed as a kind of Wheeler-deWitt equation
if we view the disk function as the Hartle-Hawking wave function. However,
$W^{(0)}_\lam(l)$ does not belong to the spectrum of $H_0(l)$ since it 
is not normalizable when one uses the measure \rf{s4}

We now introduce creation and
annihilation operators $a_n^\dg$ and $a_n$ corresponding to these states,
acting on the Fock-vacuum $\vac$ and satisfying $[a_n,a^\dg_m]=\del_{n,m}$. 
We define
\beq\label{s5}
\Psi(l) = \sum_n a_n \psi_n(l),~~~~\Psi^\dg(l) = \sum_n a_n^\dg \psi^*_n(l),
\eeq
and from the orthonormality of the eigenfunctions with respect to 
the measure $dl/l$ we recover \rf{s1}. The ``second-quantized'' Hamiltonian is
\beq\label{s6}
\hH_0 = \int_0^\infty \dll \; \Psi^\dg (l) H_0(l) \Psi(l),
\eeq
and the propagator $\tG_\lam (l_1,l_2;t)$ is now obtained as
\beq\label{s7}
\tG_\lam^{(0)} (l_1,l_2;t) = \cav \Psi(l_2) \e^{-t \hH_0} \Psi^\dg(l_1) \vac.
\eeq 

\begin{figure}[t]
\centerline{\scalebox{0.45}{\rotatebox{0}{\includegraphics{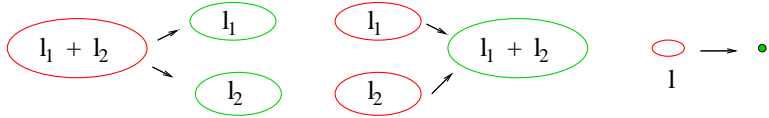}}}}
\caption[fig3]{{\small Graphical illustration of the various terms 
in eq.\ \rf{s8}.  
}}
\label{sft-fig}
\end{figure}

While this is  trivial, the advantage of the formalism
is that it automatically takes care of symmetry factors (like in the 
many-body applications in statistical field theory) both when many 
spatial universes are at play and when they are 
joining and splitting. We can follow
\cite{sft} and define the following Hamiltonian, describing the 
interaction between spatial universes:
\bea\label{s8}
\hH = \hH_0 &&-~ g_s 
\int dl_1 \int dl_2 \Psi^\dg(l_1)\Psi^\dg(l_2)\Psi(l_1+l_2)
\\ && - \a g_s\int dl_1 \int dl_2 \Psi^\dg(l_1+l_2)\Psi(l_2)\Psi(l_1)
-\int \dll \; \rho(l) \Psi(l), \nonumber
\eea
where the different terms of the Hamiltonian are illustrated in 
Fig.\ \ref{sft-fig}.
Here $g_s$ is the coupling constant we have already 
encountered in Sec.\ \ref{cap1}
of mass dimension 3. The factor $\a$ is just inserted to be 
able the identify the action of the two $g_s$-terms in \rf{s8} when
expanding in powers of $g_s$. We will think of $\a=1$ unless explicitly
stated differently.   
When $\a=1$ $\hH$ is hermitian except for the presence of the tadpole term.
It tells us that universes can vanish, but not be created from nothing.
The meaning of the two interaction terms is as follows: the first term
replaces a universe of length $l_1+l_2$ with two
universes of length $l_1$ and $l_2$. This is one of the 
processes shown in Fig.\ \ref{sft-fig}. The second term represents
the opposite process where two spatial universes merge into one,
i.e.\  the time-reversed picture. The coupling constant $g_s$ clearly 
appears as a kind of string coupling constant: one factor $g_s$ for 
splitting spatial universes, one factor $g_s$ for merging spatial
universes and thus a factor $g^2_s$ when the 
space-time topology changes, but there is also factors for 
branching alone. This is only compatible with an Euclidean sft-picture
if we associate a puncture (and thus a topology change) with the
vanishing of a baby universe. As discussed above this is indeed not
unnatural from a Lorentzian point of view. From this point of view
the appearance of a tadpole term is more 
natural in the CDT framework than in the original Euclidean 
framework in \cite{sft}. The tadpole 
term is a formal realization of this puncture ``process'', where the 
light-cone becomes degenerate. 

In principle we can now calculate the process where we start
out with $m$ spatial universes at time 0 and end with $n$ universes
at time $t$, represented as
\beq\label{s100}
G_{\lam,g_s}(l_1,..,l_m;l'_1,..,l'_n;t) =
\cav \Psi(l'_1)\ldots \Psi(l'_n) \; 
e^{-t\hH}\Psi^\dg(l_1)\ldots \Psi^\dg(l_m)\vac.
\eeq

\subsection{Dyson-Schwinger equations}\label{ds}

The disk amplitude is one of a set of functions for which 
it is possible to derive Dyson-Schwinger equations (DSE).
The disk amplitude is characterized by the fact that at $t=0$ we
have a spatial universe of some length, and at some point it vanishes
in the "vacuum".   
Let us consider the more general situation where a set of spatial universes of some 
lengths $l_i$ exists at time $t=0$ and where the universes vanish at later times.
Define the generating function:
\beq\label{ds1}
Z(J)= \lim_{t\to \infty} \cav \e^{-t \hH} \; \e^{\int dl \, J(l) \Psi^\dg(l)}\vac.
\eeq
Notice that if the tadpole term had not been present in $\hH$ $Z(J)$ 
would trivially be equal 1. We have 
\beq\label{ds2}
\lim_{t \to \infty}  \cav \e^{-t \hH} \;\Psi^\dg(l_1)\cdots \Psi^\dg(l_n)\vac = 
\left.\frac{\del^n Z(J)}{\del J(l_1)\cdots \del J(l_n)}\right|_{J=0}.
\eeq
$Z(J)$ is the generating functional for universes that disappear in the
vacuum. We now have
\beq\label{ds3}
0= \lim_{t\to \infty}\left[\frac{\prt}{\prt t}\; 
\cav \e^{-t\hH} \; \e^{\int dl \, J(l) \Psi^\dg(l)}\vac =
-\cav \e^{-t\hH} \; 
\hH\;\e^{\int dl \, J(l) \Psi^\dg(l)}\vac\right].
\eeq
Commuting the $\Psi(l)$'s in $\hH$ past the source term effectively replaces
these operators by $l J(l)$, after which they can be moved to the left of 
any $\Psi^\dg (l)$ and outside  $\cav$. 
After that the remaining $\Psi^\dg(l)$'s in $\hH$ can
be replaced by $\del/\del J(l)$ and also moved outside
$\cav$, leaving us with a integro-differential operator acting on $Z(J)$:
\beq\label{ds4}
0= \int_0^\infty dl \, J(l) \,O \left(l,J,\frac{\del}{\del J}\right)Z(J)
\eeq
where
\bea\label{ds5}
\lefteqn{\hspace{-1cm}O \left(l,J,\frac{\del}{\del J}\right)= H_0(l) \frac{\del}{\del J(l)} -
\del (l)} \\
&& -g_s l \int_0^l dl'\frac{\del^2}{\del J(l')\del J(l-l')}
-\a g_s l\int_0^\infty dl' l'J(l')\frac{\del}{\del J(l+l')}\nonumber
\eea

$Z(J)$ is a generating functional 
which also includes totally disconnected universes which 
never ``interact'' with each other.  The generating 
functional for connected universes is obtained in the standard
way from field theory by taking the logarithm of $Z(J)$. Thus 
we write:
\beq\label{ds6}
F(J) = \log Z(J),
\eeq
and we have
\beq\label{ds7}
\lim_{t \to \infty} \cav \e^{-t\hH} \Psi^\dg(l_1)\cdots \Psi^\dg(l_n)\vac_{con} =
\left.\frac{\del^n F(J)}{\del J(l_1) \cdots \del J(l_n)}\right|_{J=0}, 
\eeq
and we can readily transfer the DSE \rf{ds4}-\rf{ds5} into an equation 
for the connected functional $F(J)$:
 \bea
0= \int_0^\infty dl \, J(l)
\left\{  H_0(l)\, \frac{\del F(J)}{\del J(l)} - \delta(l) 
 -g_s l \int_0^l dl'\;\frac{\del^2 F(J)}{\del J(l')\del J(l-l')} \right. 
\nonumber\\
\left. -g_s l \int_0^l dl'
\frac{\del F(J)}{\del J(l')}\frac{\del F(J)}{\del J(l-l')}
-\a g_s l\int_0^\infty dl' l'J(l')\frac{\del F(J)}{\del J(l+l')}\right\}.
\label{ds9}
\eea
From eq.\ \rf{ds9} one obtains the DSE by differentiating 
\rf{ds9} after $J(l)$ a number of times and then taking $J(l)=0$.

\subsection{Application of the DSE}\label{application}

Let us introduce the notation 
\beq\label{ds10}
w(l_1,\ldots,l_n) \equiv 
\left.\frac{\del^n F(J)}{\del J(l_1) \cdots \del J(l_n)}\right|_{J=0}
\eeq
as well as the Laplace transform $ \tw(x_1,\ldots,x_n)$  (like in \rf{genfun1}).
Let us differentiate eq.\ \rf{ds9} after $J(l)$ one and two
times, then take $J(l)=0$
and Laplace transform the obtained equations. We obtain the 
following equations (where $H_0(x)f(x) = \prt_x [(x^2-\lam) f(x)]$):
\bea\label{ds13}
0&=&H_0(x)\tw(x) -1 +
g_s\prt_x \Big(  \tw(x,x) +  \tw(x)\tw(x)\Big),\\
&& ~ \nonumber\\
0&=&(H_0(x)+H_0(y))\tw(x,y) +g_s\prt_x \tw(x,x,y)+
g_s\prt_y  \tw(x,y,y) \label{ds15}\\
&& +2g_s\left(\prt_x [\tw(x)\tw(x,y)] \pl \prt_y[ \tw(y) \tw(x,y)]\right) +2\a g_s\prt_x\prt_y \Big(\frac{\tw(x)\mi \tw(y)}{x-y}\Big)\nonumber
\eea
The structure of the DSE for an increasing number of 
arguments is hopefully  clear (see \cite{cdt-sft} for details).

We can solve the DSE iteratively. For this purpose let us introduce 
the expansion of $\tw(x_1,\ldots,x_n)$
in terms of the coupling constants $g_s$ and $\a$:
\beq\label{ds12}
\tw(x_1,\ldots,x_n) = \sum_{k=n-1}^\infty \a^k\sum_{m=k-1}^\infty g_s^m \; 
\tw(x_1,\ldots,x_n;m,k).
\eeq
The amplitude $\tw(x_1,\ldots,x_n)$ starts with the 
power $(\a g_s)^{n-1}$ since we have to perform $n$ mergings 
during the time evolution in order to create a connected geometry
if we begin with $n$ separated spatial loops. Thus one can 
find the lowest order contribution to $\tw(x_1)$ from \rf{ds13}, use that
to find the lowest order contribution to $\tw(x_1,x_2)$ from \rf{ds15}, etc. 
Returning to eq.\ \rf{ds13}
we can use the lowest order expression for $\tw(x_1,x_2)$ to find the 
next order correction to $\tw(x_1)$, etc. 

As mentioned above the amplitude $\tw(x_1,\ldots,x_n)$ starts with the 
power $(\a g_s)^{n-1}$ coming from merging the
$n$ disconnected spatial universes. The rest of the powers
of $\a g_s$ will result in a topology change of the resulting, connected 
worldsheet. From an Euclidean point of view it is thus more appropriate
to reorganize the series as follows
\bea\label{ds12a}
&&\tw(x_1,\ldots,x_n) = (\a g_s)^{n-1}
\sum_{h=0}^\infty (\a g^2_s)^h \tw_h(x_1,\ldots,x_n)\\
&&\tw_h(x_1,\ldots,x_n) = \sum_{j=0}^\infty g_s^j 
 \tw(x_1,\ldots,x_n;n\mi 1\pl 2h \pl j,n\mi 1\pl h)\label{ds12b}
\eea
and aim for a topological expansion in $\a g^2_s$, at each order
solving for all possible baby-universe creations which at some 
point will vanish into the vacuum. Thus $\tw_h(x_1,\ldots,x_n)$ 
will be a function of $g_s$ although we do not write it explicitly.
The DSE allow us to obtain the topological expansion 
iteratively, much the same way we already did as a power expansion 
in $g_s$.

\section{The matrix model, once again}\label{matrix2}

Let us consider our $N\times N$ Hermitian matrix with the 
qubic potential \rf{3.26} and define the observable 
\beq\label{yy11}
\tW(x_1,\ldots,x_n)_d = N^{n-2} 
\left\la (\Tr \left(\frac{1}{x_1-M}\right)\cdots 
\left(\tr \frac{1}{x_1-M}\right)
\right\ra,
\eeq
where the subscript $d$ refers to the fact that the correlator
will contain disconnected parts. We denote the connected
part of the correlator by $\tW(x_1\ldots,x_n)$.
 It is standard matrix model 
technology to find the matrix model DSEs for  $\tW(x_1\ldots,x_n)$. 
We refer to \cite{davidloop,ajm,am,ackm} for details. 
{\it One obtains precisely the same set of 
coupled equations as  \rf{ds13}-\rf{ds15}  if we identify}:
\beq\label{N}
\a = \frac{1}{N^2},
\eeq
and the discussion surrounding the expansion \rf{ds12a} is 
nothing but the standard discussion of the large $N$ expansion  
of the multi-loop correlators (see for instance \cite{ackm} or 
the more recent papers \cite{eynard,ce,eo1}).
Thus we conclude that there is a perturbative agreement 
between the matrix model \rf{3.26} and the CDT SFT in the 
sense that perturbatively:
\beq\label{perturbative}
\tW(x_1,\ldots,x_n) = \tw(x_1,\ldots,x_n).
\eeq 
In practice the SFT {\it is}  only defined perturbatively, although
in principle we have available the string field Hamiltonian. However,
we can now use the matrix model to extract non-pertubative 
information. The identification of the matrix model 
and the CDT SFT DSEs were based on \rf{N}, but in the  SFT
we are interested in $\a = 1$, i.e.\ formally in $N=1$, in which 
case the matrix integrals reduce to ordinary integrals.
This means that we will consider the entire sum 
over topologies ``in one go":
\beq\label{3.1}
Z(g,\lam) = \int \d m \; 
\exp \left[ -\frac{1}{g_s} 
\left( \lam m - \frac{1}{3}\; m^3\right)\right],
\eeq 
while the observables \rf{yy11} can be written as 
\beq\label{3.2}
W_d(x_1,\ldots,x_n) = \frac{1}{Z(g_s,\lam)} 
\int \d m\; \frac{\exp \left[ -\frac{1}{g_s} 
\left( \lam m - \frac{1}{3}\; m^3\right)\right]}{(x_1-m)\cdots (x_n-m)}.
\eeq
These integrals should be understood as formal power series
in the dimensionless variable $\kp$ defined by eq.\ \rf{kappa}.
Any choice of an integration contour which makes the integral well 
defined and reproduces the formal power series is a potential
nonperturbative definition of these observables. However, different
contours might produce different nonperturbative contributions
(i.e.\ which cannot be expanded in powers of 
$t$), and there may even be nonperturbative contributions 
which are not captured by any choice of integration contour. 
As usual in such situations, additional
physics input is needed to fix these contributions.

To illustrate the point, 
let us start by evaluating the partition function given in 
\rf{3.1}. We have to decide on an integration path in the 
complex plane in order to define the integral. One possibility is to take a 
path along the negative 
axis and then along either the positive or the negative imaginary 
axis. The corresponding integrals are 
\beq\label{3.2a}
Z(g_s,\lam)= \sqrt{\lam}\; \kp^{1/3} F_{\pm} (\kp^{-2/3}),~~~
F_{\pm} (\kp^{-2/3}) =2
\pi \; e^{\pm i\pi/6}{\rm Ai}(\kp^{-2/3}\e^{\pm 2\pi i/3}),
\eeq
where Ai denotes the Airy function. Both $F_\pm$ 
have the same asymptotic expansion
in $\kp$, with positive coefficients. Had we chosen the integration path 
entirely along the imaginary axis we would have obtained ($2\pi i$ times)
${\rm Ai}(\kp^{-2/3})$, but this has an asymptotic expansion 
in $\kp$ with coefficients of oscillating sign, which is at odds with its
interpretation as a probability amplitude. In the notation of \cite{as} we have
\beq\label{3.2b}
F_{\pm}(z) = \pi \Big({\rm Bi}(z) \pm i {\rm Ai}(z)\Big),
\eeq 
from which one deduces immediately 
that the functions $F_{\pm}(\kp^{-2/3})$ are not real.
However, since ${\rm Bi}(\kp^{-2/3})$ grows like 
$e^{\frac{2}{3\kp}}$ for small $\kp$ while ${\rm Ai}(\kp^{-2/3})$ 
falls off like $e^{-\frac{2}{3\kp}}$, 
their imaginary parts are exponentially small 
in $1/\kp$ compared to the real part, and therefore do not contribute to
the asymptotic expansion in $\kp$.
An obvious way to {\it define} a partition 
function which is real and shares the
same asymptotic expansion is by symmetrization,
\beq
\oh (F_+ +F_-)\equiv \pi {\rm Bi}.
\eeq
The situation parallels the one encountered in the double scaling limit of the 
``old'' matrix model \cite{david-x}, and discussed in detail in \cite{marino},
but is less complicated. We will return to a 
discussion of this in the next section. 

Presently, let us collectively denote by $F(z)$ any of the functions 
$F_{\pm}(z)$ or $\pi {\rm Bi}(z)$, leading to the
tentative identification
\beq\label{3.3}
Z(g_s,\lam) = \sqrt{\lam}\; \kp^{1/3} \, 
F\Big(\kp^{-2/3}\Big),~~~~F''(z) = z F(z),
\eeq 
where we have included the differential equation 
satisfied by the Airy functions for
later reference. In preparation for the computation of the 
observables $\tW_d(x_1,\ldots,x_n)$ we introduce the
dimensionless variables
\beq\label{3.4}
x= \tx\,\sqrt{\lam},~~m= g_s^{1/3} \b,~~~~~
\tW_d(x_1,\ldots,x_n) = \lam^{-n/2} \tw_d(\tx_1,\ldots,\tx_n).
\eeq
Assuming $\tx_k > 0$, we can write
\beq\label{3.5}
\frac{1}{\tx-\kp^{1/3}\,\b}
= \int_0^{\infty} \d\a
\; \exp\left[-\left(\tx-\kp^{1/3}\b\right)\;\a\right].
\eeq
We can use this identity to re-express the pole terms 
in eq.\ \rf{3.2} to obtain the 
integral representation 
\beq\label{3.6}
\tw_d(\tx_1,\ldots,\tx_n) =  
\int_0^{\infty}\prod_{i=1}^n \d \a_i
\; \e^{-(\tx_1\a_1+\cdots +\tx_n\a_n)}\; 
\frac{F\Big(\kp^{-\frac{2}{3}}-
\kp^{\frac{1}{3}}\sum_{i=1}^n\a_i\Big)}{F\Big(\kp^{-\frac{2}{3}}\Big)}
\eeq
for the amplitude with dimensionless arguments.
By an inverse Laplace transformation we thus obtain:
\beq\label{3.7}
W_d(l_1,\ldots,l_n) = 
\frac{F(\kp^{-2/3}-\kp^{1/3}\sqrt{\lam}\,(l_1+\cdots+l_n))}{F(\kp^{-2/3})}.
\eeq
For the special case $n=1$ we find
\beq\label{3.8}
W(l) = \frac{F(\kp^{-2/3}-\kp^{1/3}\sqrt{\lam}\,l)}{F(\kp^{-2/3})}
\eeq
for the disc amplitude, together with the remarkable relation
\beq\label{3.8a}
W_d(l_1,\ldots,l_n)=W(l_1+\cdots+l_n).
\eeq
By Laplace transformation this formula implies the relation
\beq\label{3.8c}
\tW_d(x_1,\ldots,x_n)= 
\sum_{i=1}^n \frac{\tW(x_i)}{\prod_{j\ne i}^n (x_j-x_i)}. 
\eeq
 From $\tW_d(x_1,\ldots,x_n)$ we can construct the connected 
 multiloop functions $\tW(x_1,\ldots,x_n)$ using standard field theory. 
 Let us remark that the asymptotic
expansion in $\kp$ of $\tW(x_1,\ldots,x_n)$ of course agrees with that obtained
by recursively solving the CDT Dyson-Schwinger equations.

\section{Relation with other models}\label{other-m}

Can we identify a know continuum conformal field theory 
coupled to 2d gravity, which leads to the functions
we have here calculated? The answer is yes. 
As a starting point note that for the disk amplitude we have
\beq\label{om1}
\tW_\lam^{(0)}(x) = \frac{1}{x+\sla} = \frac{1}{x} -\frac{\sla}{x^2} + \cdots,
\eeq
from which we conclude that the susceptibility exponent 
$\g=1/2$ (the lowest non-analytic power of $\lam$ in the disk 
amplitude is $\lam^{1-\g}$).
$\g=1/2$ is the generic value of the susceptibility exponent for 
the so-called branched polymers and the way branched polymers
enters into the game of 2d gravity coupled to conformal field theories
is as follows: recall that the $(2,2m-1)$, $m=2,3,\ldots$, minimal conformal 
field theories coupled to 2d Euclidean quantum gravity can be described as 
double-scaling limits of one-matrix models with certain
fine-tuned matrix potentials of order at least $m+1$. Formally,
the case $m=1$, which corresponds to a somewhat degenerate $(2,1)$
conformal field theory with central charge $c= -2$ (which when
coupled to 2d gravity is called {\it topological gravity}), is then described
by a special double-scaling limit of the purely Gaussian matrix model
(see the review \cite{ginsparg}). In this double-scaling limit one obtains for
the so-called FZZT brane precisely the Airy function, see 
\cite{seiberg,rastelli,marino,aj} for recent discussions. For this model 
$\g =-1$. While it is possible to describe 2d topological quantum gravity  
by a double-scaling limit of the Gaussian matrix model, the 
most natural geometric interpretation of the Gaussian matrix model
is in terms of branched polymers, in the sense that the integral
\beq\label{6.3}
\frac{ \int \d M \; \tr\, M^{2n} \;\e^{-\oh \tr M^2}}{\int \d M \; 
 \e^{-\oh \tr M^2}}
\eeq
can be thought of as the gluing of a boundary of length $n$ into a double-line branched 
polymer of length $n$. Since the branched polymers are also allowed to form closed
loops, their partition function contains a sum over topologies `en miniature', and
one can indeed define a double-scaling limit of the model.
When solving for the partition function in this limit, one obtains 
precisely our $Z(g_s,\lam)$ of eq.\ \rf{3.3}! (see \cite{jk}, where this 
remarkable result was first proved,  for details). It does not imply that 
the generalized CDT model is just branched polymers, it is much richer since
it has a 2d surface representation and many more observables, but 
the non-perturbative branching process is clearly that of branched 
polymers.  The relation between
the generalized CDT model  and the $c=-2$ model is as follows:
we have the "right" branch of Liouville theory with susceptibility exponent 
$\g_-$ (= -1) and a "wrong" branch with susceptibility exponent $\g_+$, related by
\beq\label{6.2}
\g_+=  -\frac{\g_-}{1-\g_-}.
\eeq
The interpretation of this $\g_+$ in relation to $\g_-$ in terms 
geometry can be found in \cite{durhuus,adj,klebanov}, and for earlier
related work see \cite{wadia,kom}. It is all related to dominance or non-dominance
of branching of baby universes.
The simplest example is precisely given by $c=-2$: topological quantum gravity 
has $\g_-=-1$ whose dual is the ``wrong'' $\g_+ = 1/2$, which happens to be the 
value occurring generically in the theory of branched polymers 
(see, e.g., \cite{ad,ajt,ajjk} for a discussion of why branched polymers
and baby universes are generic and even dominant in many situations in non-critical
string theory and even in higher dimensional quantum gravity).

\section{Stochastic quantization}\label{stochastic}

It is a most remarkable fact that the above mentioned result can all 
be understood as a result of stochastic  quantization of {\it space}. In this
picture time becomes the {\it stochastic time} related with the branching 
of space into baby universes and the original CDT model described in 
Sec.\ \ref{cdt} becomes the classical limit where no stochastic 
processes are present \cite{sto-alwz}.

Recall the Langevin stochastic differential equation for a single variable
$x$ (see, for example, \cite{zinn1,chai}).
\beq\label{x2.1}
\dot{x}^{(\n)}(t) = - f\Big(x^{(\n)}(t)\Big) + \sOm \;\n(t),
\eeq
where the dot denotes differentiation with respect to stochastic time $t$,
$\n(t)$ is a Gaussian white-noise term of unit width 
and $f(x)$ a dissipative drift force:
\beq\label{x2.2}
f(x) =  \frac{\prt S(x)}{\prt x}
\eeq
 The noise term creates a probability distribution of $x(t)$, reflecting
the assumed stochastic nature of the noise term, with an associated 
probability distribution 
\beq\label{x2.3}
P(x,x_0;t) = \lla \del (x-x^{(\n)}(t;x_0))\rra_\n,
\eeq
where the expectation value refers to an average over the Gaussian noise.
$P(x,x_0;t)$ satisfies the Fokker-Planck equation
\beq\label{x2.5}
\frac{\prt P(x,x_0;t)}{\prt t} = 
 \frac{\prt}{\prt x}\left( \oh\Om \frac{\prt P(x,x_0;t) }{\prt x} + 
f(x) P(x,x_0;t)\right).
\eeq
This is an imaginary-time Schr\"{o}dinger equation, with
$\sOm$ playing a role similar to $\hbar$. It enables us to write
$P$ as a propagator for a Hamiltonian operator $\hH$,
\beq\label{x2.6}
P(x,x_0;t) = \la x | \e^{-t\hH}|x_0\ra,~~~\hH= \oh \Om \hp^2 +i \hp \, f(\hx),
\eeq
with initial condition $x(t=0) = x_0$, and $\hp = -i \prt_x$. It follows that
by defining
\beq\label{x2.6a}
\tG(x_0,x;t) \equiv \frac{\prt}{\prt x_0}\, P(x,x_0;t) 
\eeq
the function $\tG(x_0,x;t)$ satisfies the differential equation
\beq\label{x2.6b} 
\frac{\prt \tG(x_0,x;t)}{\prt t} = 
 \frac{\prt}{\prt x_0}\left( \oh \,\Om\, \frac{\prt \tG(x_0,x;t)}{\prt x_0} - 
f(x_0)\;\tG(x_0,x;t)\right).
\eeq

Omitting the noise term corresponds
to taking the limit $\Om \to 0$. One can then drop the functional average
over the noise in \rf{x2.3} to obtain
\beq\label{x2.21}
P_{cl}(x,x_0;t) = \del(x-x(t,x_0)),~~~~\tG_{cl}(x_0,x;t) = 
\frac{\prt}{\prt x_0} \del(x-x(t,x_0)).
\eeq
It is readily verified that these functions satisfy 
eqs.\ \rf{x2.5} and \rf{x2.6b} with $\Om =0$. Thus we 
have for $S(x) = -\lam x +x^3/3$:
\beq\label{x2.22} 
\frac{\prt \tG_{cl}(x_0,x;t)}{\prt t} = 
 \frac{\prt}{\prt x_0} \Big( (\lam - x_0^2 )\, \tG_{cl}(x_0,x;t)\Big).
\eeq  

Comparing now eqs.\ \rf{cdt32} and \rf{x2.22}, 
we see that we can formally re-interpret 
$\tG_\lam^{(0)}(x_0,x;t)$ -- an amplitude 
obtained by nonperturbatively quantizing Lorentzian
pure gravity in two dimensions -- 
as the ``{\it classical} probability'' $\tG_{cl}(x_0,x;t)$ corresponding
to the action $S(x) = -\lam x +x^3/3$ of a zero-dimensional system
in the context of stochastic quantization, only is the boundary condition
different, since in the case of CDT $x$ is not an ordinary 
real variable, but the cosmological constant. The correct boundary conditions
are thus the ones stated in eqs.\ \rf{cdt78}--\rf{cdt79}

Stochastic quantization of the system amounts to replacing
\beq\label{y2.5}
\tG_\lam^{(0)} (x_0,x;t) \to \tG(x_0,x;t),
\eeq
where $\tG(x_0,x;t)$ satisfies the differential equation
corresponding to eq.\ \rf{x2.6b}, namely,
\beq\label{y2.6}
\frac{\prt \tG(x_0,x;t)}{\prt t} = 
 \frac{\prt}{\prt x_0}\left( g_s \frac{\prt}{\prt x_0} + \lam-x_0^2\right) 
\tG(x_0,x;t).
\eeq
We have introduced the
parameter $g_s:=\Om/2$, which will allow us to reproduce 
the matrix model and SFT results reported above.   

 A neat geometric interpretation
of how stochastic quantization can capture topologically nontrivial
amplitudes has been given in \cite{sto-kawai}. Applied to the present
case, we can view 
the propagation in stochastic time $t$ for a given noise term $\n(t)$ 
as classical in the sense that solving the 
Langevin equation \rf{x2.1} for $x^{(\n)}(t)$ iteratively gives
precisely the tree diagrams with one external leg 
corresponding to the action $S(x)$ (and including the derivative
term $\dot x^{(\n)}(t)$), with 
the noise term acting as a source term. Performing the functional 
integration over the Gaussian noise term corresponds to integrating out the 
sources and creating loops, or, if we have several independent trees,
to merging these trees and creating diagrams with several external legs.
If the dynamics of the quantum states of the spatial universe 
takes place via the strictly causal CDT-propagator $\hG_0 = \e^{-t \hH_0}$, 
a single spatial universe of length $l$ 
cannot split into two spatial universes. Similarly, no two spatial universes are 
allowed to merge as a function of stochastic time.
However, introducing the noise term {\it and} subsequently
performing a functional
integration over it makes these processes possible. 
This explains how the stochastic quantization can automatically generate
the amplitudes which are introduced by hand
in a string field theory, be it of Euclidean character as described
in \cite{sto-kawai}, or within the framework of CDT.

What is new in the CDT string field theory considered
here is that we can use the corresponding 
stochastic field theory to solve the model.
since we arrive at 
closed equations valid to all orders in the genus expansion.
Let us translate equations \rf{y2.6} to $l$-space 
\beq\label{3.11}
 \frac{\prt G(l_0,l;t)}{\prt t} =-H(l_0)\, G(l_0,l;t),
\eeq
where the {\it extended} Hamiltonian
\beq\label{3.12}
H(l) = -l \frac{\prt^2}{\prt l^2} +\lam l - g_s l^2 = H_0(l)-g_s l^2
\eeq
now has an extra potential term coming from the inclusion of branching 
points compared to the Hamiltonian $H_0(l)$ defined in \rf{35b}. 
It is truely remarkable that all branching and joining is contained in 
this simple extra term. Formally  $H(l)$ is a well-defined Hermitian operator with respect to the measure \rf{s4} (we will discuss some subtleties in the 
next section). 

We can now write down the generalization of  Wheeler-deWitt equation
\rf{s4} for the disk amplitude 
\beq\label{3.11a}
H(l) W(l)=0.
\eeq
Contrary to $W_\lam^{(0)}(l)$ appearing in \rf{s4},  $W(l)$ contains
all branchings and all topology changes, and the solution is precisely 
\rf{3.8}! This justifies the choice $g_s=\Om/2$ mentioned above.
Recall that  $E=0$ does not belong to the spectrum of 
$H_0(l)$ since $W_0(l)$ is not integrable at zero with respect to the measure
\rf{s4}. Exactly the same is true for the extended Hamiltonian $H(l)$ and 
the corresponding Hartle-Hawking amplitude $W(l)$

We have also as a generalization of \rf{ham}  that
\beq\label{3.13}
G(l_0,l;t) = \la l | e^{-t H(l)}|l_0\ra
\eeq
describes the nonperturbative propagation of a spatial loop 
of length $l_0$ to a spatial loop of length $l$ in proper 
(or stochastic) time $t$, now including the summation over all genera.

\section{The extended Hamiltonian}\label{hamiltonian}

In order to analyze the spectrum of $H(l)$,
it is convenient to put the differential operator into standard form.
After a change of variables
\beq\label{4.3}
l= \oh z^2,~~~~~\psi(l) = \sqrt{z} \phi(z),
\eeq
the eigenvalue equation becomes
\beq\label{4.4}
H(z)\phi(z) = E \phi(z),~~~~H(z) = -\oh \frac{\d^2}{\d z^2} 
+\oh \lam z^2 + \frac{3}{8z^2}-\frac{g_s}{4} z^4.
\eeq 
This shows that the potential is unbounded from below, but 
such that the eigenvalue spectrum is still discrete: whenever
the potential is unbounded below with fall-off faster than $- z^2$, the spectrum is discrete, reflecting the fact that the classical escape time to infinity is finite
(see \cite{ak} for a detailed discussion relevant to the present situation).
For small $g_s$,  there is a large barrier of height $\lam^2/(2g_s)$ 
separating the unbounded region 
for $l > \lam/g_s$ from the region $0 \leq l \leq \lam/(2g_s)$ where the 
potential grows. This situation is perfectly suited to applying a standard WKB 
analysis. For energies less than $\lam^2/(2g_s)$, the eigenfunctions
\rf{s4a} of $H_0(l)$ will be good approximations to those of $\hH(l)$. 
However, when $l > \lam/g_s$ the exponential fall-off of $\psi_n^{(0)}(l)$
will be replaced by an oscillatory behaviour, with the wave function falling 
off only like $1/l^{1/4}$. The corresponding $\psi_n(l)$ is still
square-integrable since we have to use the measure \rf{s4}.
For energies larger than  $\lam^2/(2g_s)$, the solutions will be 
entirely oscillatory, but still square-integrable.

Thus a somewhat drastic change has occurred in the quantum behaviour of the one-dimensional universe as a consequence of allowing topology changes.
In the original, strictly causal quantum gravity model 
an eigenstate $\psi_n^{(0)}(l)$ of the spatial universe had an average size
of order $1/\sqrt{\lam}$.
However, allowing for branching and topology change, the average size of the universe is now infinite!

As discussed in \cite{ak}, Hamiltonians with unbounded potentials like 
\rf{4.4} have a one-parameter family of selfadjoint extensions and we 
still have to choose one of those such that the spectrum of $H(l)$ can be determined unambiguously.
One way of doing this is to appeal again to stochastic quantization, 
following the strategy used by Greensite and Halpern \cite{gh}, 
which was applied to the double-scaling limit of matrix models in \cite{ag,ag1,ak}.
The Hamiltonian \rf{x2.6} corresponding to the Fokker-Planck equation \rf{y2.6}, namely,
\beq\label{4.5}
H(x)\psi(x) = -g_s \frac{\d^2 \psi(x)}{\d x^2} +\frac{\d}{\d x}
\left(\frac{\d S(x)}{\d x}\, \psi(x) \right),~~~~ 
S(x) = \left(\frac{x^3}{3}-\lam\,x\right),
\eeq  
is not Hermitian if we view $x$ as an ordinary real 
variable and wave functions $\psi(x)$ as endowed with the standard scalar
product on the real line. However, by a similarity transformation one can transform $H(x)$
to a new operator 
\beq\label{4.6}
\tH(x) = \e^{-S(x)/2g_s}H(x) \, \e^{S(x)/2g_s};~~~
\tilde{\psi}(x) = \e^{-S(x)/2g_s}\psi(x),
\eeq 
which {\it is} Hermitian on $L^2(R,dx)$.
We have 
\beq\label{4.7}
\tH(x)= -g_s\frac{\d^2}{\d x^2} +
\left(\frac{1}{4g_s} \left(\frac{\d S(x)}{\d x}\right)^2+
\oh \frac{\d^2 S(x)}{\d x^2}\right),
\eeq
which after substitution of the explicit form of the action becomes
\beq\label{4.7a}
\tH(x)=-g_s \frac{\d^2}{\d x^2} +V(x),~~~~V(x)= \frac{1}{4g_s} (\lam -x^2)^2+ x.
\eeq
The fact that one can write 
\beq\label{4.8}
\tH(x)=  R^{\dg}R,~~~~
R=-\sqrt{g_s}\frac{\d}{\d x} +\frac{1}{2\sqrt{g_s}}\frac{\d S(x)}{\d x}
\eeq
implies that the spectrum of $\tH(x)$ is positive, discrete and 
unambiguous. We conclude that the formalism of stochastic quantization
has provided us with a nonperturbative definition of the CDT 
string field theory.

\subsection*{Acknowledgments}
JA, RL, WW and SZ acknowledge support by
ENRAGE (European Network on
Random Geometry), a Marie Curie Research Training Network, 
contract MRTN-CT-2004-005616, and 
RL acknowledges support by the Netherlands
Organisation for Scientific Research (NWO) under their VICI
program.
SZ thanks the Department of Statistics at Sao Paulo University for kind
hospitality and acknowledges financial support of the ISAC program,
Erasmus Mundus.

\end{document}